
%
%
  \font\twelverm=cmr10 scaled 1200       \font\twelvei=cmmi10 scaled 1200
  \font\twelvesy=cmsy10 scaled 1200      \font\twelveex=cmex10 scaled 1200
  \font\twelvebf=cmbx10 scaled 1200      \font\twelvesl=cmsl10 scaled 1200
  \font\twelvett=cmtt10 scaled 1200      \font\twelveit=cmti10 scaled 1200
  \font\twelvemib=cmmib10 scaled 1200
  \font\elevenmib=cmmib10 scaled 1095
  \font\tenmib=cmmib10
  \font\eightmib=cmmib10 scaled 800
  
\font\elevenrm=cmr10 scaled 1095    \font\eleveni=cmmi10 scaled 1095
\font\elevensy=cmsy10 scaled 1095

%
%

\font\seventeeni=cmmi10 scaled \magstep3

\font\seventeensy=cmsy10 scaled \magstep3

\font\seventeenmib=cmmib10 scaled \magstep3

\newfam\cpfam%



\skewchar\eleveni='177   \skewchar\elevensy='60
\skewchar\elevenmib='177  \skewchar\seventeensy='60
\skewchar\seventeenmib='177
\skewchar\seventeeni='177

\newfam\mibfam%


  \skewchar\twelvei='177   \skewchar\twelvesy='60
  \skewchar\twelvemib='177
%
%
\def\twelvepoint{\normalbaselineskip=12.4pt
  \abovedisplayskip 12.4pt plus 3pt minus 9pt
  \belowdisplayskip 12.4pt plus 3pt minus 9pt
  \abovedisplayshortskip 0pt plus 3pt
  \belowdisplayshortskip 7.2pt plus 3pt minus 4pt
  \smallskipamount=3.6pt plus 1.2pt minus 1.2pt
  \medskipamount=7.2pt plus 2.4pt minus 2.4pt
  \bigskipamount=14.4pt plus 4.8pt minus 4.8pt
  \def\rm{\fam0\twelverm}          \def\it{\fam\itfam\twelveit}%
  \def\sl{\fam\slfam\twelvesl}     \def\bf{\fam\bffam\twelvebf}%
  \def\mit{\fam 1}                 \def\cal{\fam 2}%
  \def\tt{\twelvett}%
  \def\mib{\fam\mibfam\twelvemib}%

  \textfont0=\twelverm   \scriptfont0=\tenrm     \scriptscriptfont0=\sevenrm
  \textfont1=\twelvei    \scriptfont1=\teni      \scriptscriptfont1=\seveni
  \textfont2=\twelvesy   \scriptfont2=\tensy     \scriptscriptfont2=\sevensy
  \textfont3=\twelveex   \scriptfont3=\twelveex  \scriptscriptfont3=\twelveex
  \textfont\itfam=\twelveit
  \textfont\slfam=\twelvesl
  \textfont\bffam=\twelvebf
  \textfont\mibfam=\twelvemib       \scriptfont\mibfam=\tenmib
                                             \scriptscriptfont\mibfam=\eightmib

  \def\xrm{\textfont0=\twelverm\scriptfont0=\tenrm
      \scriptscriptfont0=\sevenrm\rm}
\normalbaselines\rm}


\mathchardef\alpha="710B
\mathchardef\beta="710C
\mathchardef\gamma="710D
\mathchardef\delta="710E
\mathchardef\epsilon="710F
\mathchardef\zeta="7110
\mathchardef\eta="7111
\mathchardef\theta="7112
\mathchardef\kappa="7114
\mathchardef\lambda="7115
\mathchardef\mu="7116
\mathchardef\nu="7117
\mathchardef\xi="7118
\mathchardef\pi="7119
\mathchardef\rho="711A
\mathchardef\sigma="711B
\mathchardef\tau="711C
\mathchardef\phi="711E
\mathchardef\chi="711F
\mathchardef\psi="7120
\mathchardef\omega="7121
\mathchardef\varepsilon="7122
\mathchardef\vartheta="7123
\mathchardef\varrho="7125
\mathchardef\varphi="7127

\def\physgreek{
\mathchardef\Gamma="7100
\mathchardef\Delta="7101
\mathchardef\Theta="7102
\mathchardef\Lambda="7103
\mathchardef\Xi="7104
\mathchardef\Pi="7105
\mathchardef\Sigma="7106
\mathchardef\Upsilon="7107
\mathchardef\Phi="7108
\mathchardef\Psi="7109
\mathchardef\Omega="710A}


\def\beginlinemode{\endmode
  \begingroup\parskip=0pt \obeylines\def\\{\par}\def\endmode{\par\endgroup}}
\def\beginparmode{\endmode
  \begingroup \def\endmode{\par\endgroup}}
\let\endmode=\par
{\obeylines\gdef\
{}}
\def\singlespace{\baselineskip=\normalbaselineskip}

\def\oneandahalfspace{\baselineskip=\normalbaselineskip
  \multiply\baselineskip by 3 \divide\baselineskip by 2}
\def\doublespace{\baselineskip=\normalbaselineskip \multiply\baselineskip by 2}

\nopagenumbers
\newcount\firstpageno
\firstpageno=2
\headline={\ifnum\pageno<\firstpageno{\hfil}\else{\hfil\elevenrm\folio\hfil}\fi}
\let\rawfootnote=\footnote             
\def\footnote#1#2{{\singlespace\parindent=0pt
\rawfootnote{#1}{#2}}}
\def\raggedcenter{\leftskip=4em plus 12em \rightskip=\leftskip
  \parindent=0pt \parfillskip=0pt \spaceskip=.3333em \xspaceskip=.5em
  \pretolerance=9999 \tolerance=9999
  \hyphenpenalty=9999 \exhyphenpenalty=9999 }
\def\dateline{\rightline{\ifcase\month\or
  January\or February\or March\or April\or May\or June\or
  July\or August\or September\or October\or November\or December\fi
  \space\number\year}}
\def\received{\vskip 3pt plus 0.2fill
 \centerline{\sl (Received\space\ifcase\month\or
  January\or February\or March\or April\or May\or June\or
  July\or August\or September\or October\or November\or December\fi
  \qquad, \number\year)}}


\hsize=6.5truein
\hoffset=0.0truein
\vsize=8.9truein
\voffset=0truein
\hfuzz=0.1pt
\vfuzz=0.1pt
\parskip=\medskipamount
\overfullrule=0pt      



\def\title                     
  {\null\vskip 3pt plus 0.1fill
   \beginlinemode \doublespace \raggedcenter \bf}

\def\author                    
  {\vskip 6pt plus 0.2fill \beginlinemode
   \singlespace \raggedcenter}

\def\affil        
  {\vskip 6pt plus 0.1fill \beginlinemode
   \oneandahalfspace \raggedcenter \it}

\def\abstract                  
  {\vskip 6pt plus 0.3fill \beginparmode
   \doublespace \narrower }

\def\summary                   
  {\vskip 3pt plus 0.3fill \beginparmode
   \doublespace \narrower SUMMARY: }

\def\pacs#1
  {\vskip 3pt plus 0.2fill PACS numbers: #1}

\def\endtitlepage              
  {\endpage                    
   \body}

\def\body                      
  {\beginparmode}              

\def\head#1{                   
  \filbreak\vskip 0.5truein    
  {\immediate\write16{#1}
   \raggedcenter \uppercase{#1}\par}
   \nobreak\vskip 0.25truein\nobreak}

%
%

%
\def\inlinerefs{
  \gdef\refto##1{ [##1]}                
\gdef\refis##1{\indent\hbox to 0pt{\hss##1.~}} 
\gdef\journal##1, ##2, ##3, 1##4##5##6{ 
    {\sl ##1~}{\bf ##2}, ##3 (1##4##5##6)}}    
\def\keywords#1
  {\vskip 3pt plus 0.2fill Keywords: #1}
\gdef\figis#1{\indent\hbox to 0pt{\hss#1.~}} 

\def\figurecaptions     
  {\head{Figure Captions}    
   \beginparmode
   \interlinepenalty=10000
   \frenchspacing \parindent=0pt \leftskip=1truecm
   \parskip=8pt plus 3pt \everypar{\hangindent=\parindent}}

%
%
\def\refto#1{$^{#1}$}          

\def\references       
  {\head{References}           
   \beginparmode
   \frenchspacing \parindent=0pt \leftskip=1truecm
   \interlinepenalty=10000
   \parskip=8pt plus 3pt \everypar{\hangindent=\parindent}}

\gdef\refis#1{\indent\hbox to 0pt{\hss#1.~}} 

\gdef\journal#1, #2, #3, 1#4#5#6{              
    {\sl #1~}{\bf #2}, #3 (1#4#5#6)}          

\def\refstylenp{               
  \gdef\refto##1{ [##1]}                               
  \gdef\refis##1{\indent\hbox to 0pt{\hss##1)~}}      
  \gdef\journal##1, ##2, ##3, ##4 {                    
     {\sl ##1~}{\bf ##2~}(##3) ##4 }}

\def\refstyleprnp{             
  \gdef\refto##1{ [##1]}                               
  \gdef\refis##1{\indent\hbox to 0pt{\hss##1)~}}      
  \gdef\journal##1, ##2, ##3, 1##4##5##6{              
    {\sl ##1~}{\bf ##2~}(1##4##5##6) ##3}}

\def\prb{\journal Phys. Rev. B, }

\def\prl{\journal Phys. Rev. Lett., }

\def\endreferences{\body}

%
%

\def\endfigurecaptions{\body}

\def\endpage                   
  {\vfill\eject}

\def\endpaper                  
  {\endmode\vfill\supereject}


\def\ref#1{Ref.[#1]}                   
\def\Ref#1{Ref.[#1]}                   

\def\Fig#1{Fig.[#1]}

\def\Equation#1{Equation [#1]}         
\def\Equations#1{Equations [#1]}       
\def\Eq#1{Eq. (#1)}                     
\def\eq#1{Eq. (#1)}                     
\def\Eqs#1{Eqs. (#1)}                   
\def\eqs#1{Eqs. (#1)}                   
\def\frac#1#2{{\textstyle{{\strut #1} \over{\strut #2}}}}

\def\sla{\raise.15ex\hbox{$/$}\kern-.57em}
\def\leaderfill{\leaders\hbox to 1em{\hss.\hss}\hfill}
\def\twiddle{\lower.9ex\rlap{$\kern-.1em\scriptstyle\sim$}}
\def\bigtwiddle{\lower1.ex\rlap{$\sim$}}
\def\gtwid{\mathrel{\raise.3ex\hbox{$>$\kern-.75em\lower1ex\hbox{$\sim$}}}}
\def\ltwid{\mathrel{\raise.3ex\hbox{$<$\kern-.75em\lower1ex\hbox{$\sim$}}}}
\def\square{\kern1pt\vbox{\hrule height 1.2pt\hbox{\vrule width 1.2pt\hskip 3pt
   \vbox{\vskip 6pt}\hskip 3pt\vrule width 0.6pt}\hrule height 0.6pt}\kern1pt}

%

%

%

%
\physgreek
%

\def\dsl{\raise.15ex\hbox{$/$}\kern-.57em\hbox{$\partial$}}
\def\nsl{\raise.15ex\hbox{$/$}\kern-.57em\hbox{$\nabla$}}
\def\gtwid{\,{\raise.3ex\hbox{$>$\kern-.75em\lower1ex\hbox{$\sim$}}}\,}
\def\ltwid{\,{\raise.3ex\hbox{$<$\kern-.75em\lower1ex\hbox{$\sim$}}}\,}
\def\undr{\raise.3ex\hbox{$\sim$\kern-.75em\lower1ex\hbox{$|\vec
x|\to\infty$}}}

\def\[{\left [}
\def\]{\right ]}
\def\({\left (}
\def\){\right )}







\def\and{a^{\phantom\dagger}}

%
\def\id{\raise.72ex\hbox{$-$}\kern-.85em\hbox{$d$}\,}

\catcode`@=11
\newcount\r@fcount \r@fcount=0
\newcount\r@fcurr
\immediate\newwrite\reffile
\newif\ifr@ffile\r@ffilefalse
\def\w@rnwrite#1{\ifr@ffile\immediate\write\reffile{#1}\fi\message{#1}}

\def\writer@f#1>>{}
\def\referencefile{
  \r@ffiletrue\immediate\openout\reffile=\jobname.ref%
  \def\writer@f##1>>{\ifr@ffile\immediate\write\reffile%
    {\noexpand\refis{##1} = \csname r@fnum##1\endcsname = %
     \expandafter\expandafter\expandafter\strip@t\expandafter%
     \meaning\csname r@ftext\csname r@fnum##1\endcsname\endcsname}\fi}%
  \def\strip@t##1>>{}}

\def\citeall#1{\xdef#1##1{#1{\noexpand\cite{##1}}}}
\def\cite#1{\each@rg\citer@nge{#1}}	

\def\each@rg#1#2{{\let\thecsname=#1\expandafter\first@rg#2,\end,}}
\def\first@rg#1,{\thecsname{#1}\apply@rg}	
\def\apply@rg#1,{\ifx\end#1\let\next=\relax
\else,\thecsname{#1}\let\next=\apply@rg\fi\next}

\def\citer@nge#1{\citedor@nge#1-\end-}	
\def\citer@ngeat#1\end-{#1}
\def\citedor@nge#1-#2-{\ifx\end#2\r@featspace#1 
  \else\citel@@p{#1}{#2}\citer@ngeat\fi}	
\def\citel@@p#1#2{\ifnum#1>#2{\errmessage{Reference range #1-#2\space is bad.}%
    \errhelp{If you cite a series of references by the notation M-N, then M and
    N must be integers, and N must be greater than or equal to M.}}\else%
 {\count0=#1\count1=#2\advance\count1
by1\relax\expandafter\r@fcite\the\count0,%
  \loop\advance\count0 by1\relax
    \ifnum\count0<\count1,\expandafter\r@fcite\the\count0,%
  \repeat}\fi}

\def\r@featspace#1#2 {\r@fcite#1#2,}	
\def\r@fcite#1,{\ifuncit@d{#1}
    \newr@f{#1}%
    \expandafter\gdef\csname r@ftext\number\r@fcount\endcsname%
                     {\message{Reference #1 to be supplied.}%
                      \writer@f#1>>#1 to be supplied.\par}%
 \fi%
 \csname r@fnum#1\endcsname}
\def\ifuncit@d#1{\expandafter\ifx\csname r@fnum#1\endcsname\relax}%
\def\newr@f#1{\global\advance\r@fcount by1%
    \expandafter\xdef\csname r@fnum#1\endcsname{\number\r@fcount}}

\let\r@fis=\refis			
\def\refis#1#2#3\par{\ifuncit@d{#1}
   \newr@f{#1}%
   \w@rnwrite{Reference #1=\number\r@fcount\space is not cited up to now.}\fi%
  \expandafter\gdef\csname r@ftext\csname r@fnum#1\endcsname\endcsname%
  {\writer@f#1>>#2#3\par}}

\def\ignoreuncited{
   \def\refis##1##2##3\par{\ifuncit@d{##1}%
     \else\expandafter\gdef\csname r@ftext\csname
r@fnum##1\endcsname\endcsname%
     {\writer@f##1>>##2##3\par}\fi}}

\def\r@ferr{\endreferences\errmessage{I was expecting to see
\noexpand\endreferences before now;  I have inserted it here.}}
\let\r@ferences=\references
\def\references{\r@ferences\def\endmode{\r@ferr\par\endgroup}}

\let\endr@ferences=\endreferences
\def\endreferences{\r@fcurr=0
  {\loop\ifnum\r@fcurr<\r@fcount
    \advance\r@fcurr by 1\relax\expandafter\r@fis\expandafter{\number\r@fcurr}%
    \csname r@ftext\number\r@fcurr\endcsname%
  \repeat}\gdef\r@ferr{}\endr@ferences}


\let\r@fend=\endpaper\gdef\endpaper{\ifr@ffile
\immediate\write16{Cross References written on []\jobname.REF.}\fi\r@fend}

\catcode`@=12

\citeall\refto		
\citeall\ref		%
\citeall\Ref		%

\catcode`@=11
\newcount\tagnumber\tagnumber=0

\immediate\newwrite\eqnfile
\newif\if@qnfile\@qnfilefalse
\def\write@qn#1{}
\def\writenew@qn#1{}
\def\w@rnwrite#1{\write@qn{#1}\message{#1}}
\def\@rrwrite#1{\write@qn{#1}\errmessage{#1}}

\def\taghead#1{\gdef\t@ghead{#1}\global\tagnumber=0}
\def\t@ghead{}

\expandafter\def\csname @qnnum-3\endcsname
  {{\t@ghead\advance\tagnumber by -3\relax\number\tagnumber}}
\expandafter\def\csname @qnnum-2\endcsname
  {{\t@ghead\advance\tagnumber by -2\relax\number\tagnumber}}
\expandafter\def\csname @qnnum-1\endcsname
  {{\t@ghead\advance\tagnumber by -1\relax\number\tagnumber}}
\expandafter\def\csname @qnnum0\endcsname
  {\t@ghead\number\tagnumber}
\expandafter\def\csname @qnnum+1\endcsname
  {{\t@ghead\advance\tagnumber by 1\relax\number\tagnumber}}
\expandafter\def\csname @qnnum+2\endcsname
  {{\t@ghead\advance\tagnumber by 2\relax\number\tagnumber}}
\expandafter\def\csname @qnnum+3\endcsname
  {{\t@ghead\advance\tagnumber by 3\relax\number\tagnumber}}

\def\equationfile{%
  \@qnfiletrue\immediate\openout\eqnfile=\jobname.eqn%
  \def\write@qn##1{\if@qnfile\immediate\write\eqnfile{##1}\fi}
  \def\writenew@qn##1{\if@qnfile\immediate\write\eqnfile
    {\noexpand\tag{##1} = (\t@ghead\number\tagnumber)}\fi}
}

\def\callall#1{\xdef#1##1{#1{\noexpand\call{##1}}}}
\def\call#1{\each@rg\callr@nge{#1}}

\def\each@rg#1#2{{\let\thecsname=#1\expandafter\first@rg#2,\end,}}
\def\first@rg#1,{\thecsname{#1}\apply@rg}
\def\apply@rg#1,{\ifx\end#1\let\next=\relax%
\else,\thecsname{#1}\let\next=\apply@rg\fi\next}

\def\callr@nge#1{\calldor@nge#1-\end-}
\def\callr@ngeat#1\end-{#1}
\def\calldor@nge#1-#2-{\ifx\end#2\@qneatspace#1 %
  \else\calll@@p{#1}{#2}\callr@ngeat\fi}
\def\calll@@p#1#2{\ifnum#1>#2{\@rrwrite{Equation range #1-#2\space is bad.}
\errhelp{If you call a series of equations by the notation M-N, then M and
N must be integers, and N must be greater than or equal to M.}}\else%
 {\count0=#1\count1=#2\advance\count1
 by1\relax\expandafter\@qncall\the\count0,%
  \loop\advance\count0 by1\relax%
    \ifnum\count0<\count1,\expandafter\@qncall\the\count0,%
  \repeat}\fi}

\def\@qneatspace#1#2 {\@qncall#1#2,}
\def\@qncall#1,{\ifunc@lled{#1}{\def\next{#1}\ifx\next\empty\else
  \w@rnwrite{Equation number \noexpand\(>>#1<<) has not been defined yet.}
  >>#1<<\fi}\else\csname @qnnum#1\endcsname\fi}

\let\eqnono=\eqno
\def\eqno(#1){\tag#1}
\def\tag#1$${\eqnono(\displayt@g#1 )$$}

\def\aligntag#1\endaligntag
  $${\gdef\tag##1\\{&(##1 )\cr}\eqalignno{#1\\}$$
  \gdef\tag##1$${\eqnono(\displayt@g##1 )$$}}

\def\eqalignno#1{\displ@y \tabskip\centering
  \halign to\displaywidth{\hfil$\displaystyle{##}$\tabskip\z@skip
    &$\displaystyle{{}##}$\hfil\tabskip\centering
    &\llap{$\displayt@gpar##$}\tabskip\z@skip\crcr
    #1\crcr}}

\def\displayt@gpar(#1){(\displayt@g#1 )}

\def\displayt@g#1 {\rm\ifunc@lled{#1}\global\advance\tagnumber by1
        {\def\next{#1}\ifx\next\empty\else\expandafter
        \xdef\csname @qnnum#1\endcsname{\t@ghead\number\tagnumber}\fi}%
  \writenew@qn{#1}\t@ghead\number\tagnumber\else
        {\edef\next{\t@ghead\number\tagnumber}%
        \expandafter\ifx\csname @qnnum#1\endcsname\next\else
        \w@rnwrite{Equation \noexpand\tag{#1} is a duplicate number.}\fi}%
  \csname @qnnum#1\endcsname\fi}

\def\ifunc@lled#1{\expandafter\ifx\csname @qnnum#1\endcsname\relax}

\let\@qnend=\end\gdef\end{\if@qnfile
\immediate\write16{Equation numbers written on []\jobname.EQN.}\fi\@qnend}

\catcode`@=12
\callall\Equation
\callall\Equations
\callall\Eq
\callall\eq
\callall\Eqs
\callall\eqs


\twelvepoint
\doublespace
\referencefile
\overfullrule=10pt

\title{ Magnetic Oscillations of a Fractional Hall Dot}
\author A.H. MacDonald$^{(a)}$
\affil Max-Planck-Institut f\"ur Festk\"orperforschung\\
D-7000 Stuttgart 80\\ Germany
\author M.D. Johnson
\affil Department of Physics\\University of Central
Florida\\Orlando FL 32816-2385

\abstract { We show that a quantum dot in the
fractional Hall regime exhibits mesoscopic magnetic oscillations
with a period which is a multiple of the period for free electrons.
Our calculations are performed for parabolic quantum dots with
hard-core electron-electron interactions and are exact
in the strong field limit for $k_B T$ smaller than the fractional
Hall gap.  Explicit expressions are given for the temperature
dependence of the amplitude of the oscillations.}

\pacs {73.20.Dx,73.20.Mf}

\endtitlepage

  Advances in nanofabrication technology have made it
possible to realize artificial systems in which
electrons are confined to a small area within a
two-dimensional electron gas.  There
has been considerable interest in the physics of electron-electron
and electron-hole interactions in these `quantum dot'
systems\refto{qdot}.
Recent experiments have demonstrated the
possibility of probing their properties in
the regimes of the integer\refto{kastner} and
fractional\refto{hansen} quantum Hall effects\refto{qhe}.
The quantum Hall effect occurs whenever a system has a
chemical potential discontinuity in the thermodynamic limit
at a magnetic field dependent density.
The fractional Hall gaps occur because of electron-electron
interactions.  Their microscopic origin
is understood in some detail and that understanding is
exploited here to discuss the
thermodynamic properties of parabolic quantum dot
systems in the fractional
quantum Hall regime.  Our calculations are exact within
the hard-core model\refto{kivtrug,haldanehc} which may be
considered as an ideal model of the fractional quantum
Hall effect.  We show that in the regime where the
electronic ground state of the dot is the Laughlin\refto{laugh}
state which occurs at filling factor $\nu =1/m$ in the bulk,
all thermodynamic properties of electrons in the dot have a component
which is periodic in the magnetic flux through the area of the
dot with period $m \Phi_0$.  (Here $\Phi_0 \equiv hc/e $ is the
electronic magnetic flux quantum.)

  We consider a system of electrons in contact with a
particle reservoir with chemical potential $\mu$
and confined to a finite area of a two
dimensional electron gas by a parabolic potential,
$V(r) = {1 \over 2} m \Omega^2 r^2 $.
We measure the chemical potential
with respect to the two-dimensional subband energy and define
a nominal radius, $R$, for the quantum dot system in terms of the
radius where the parabolic confining potential equals the amount
by which the chemical potential exceeds the
quantized kinetic energy of the lowest Landau level;
$ {1 \over 2} m \Omega^2 R^2 \equiv \mu - hbar
\omega_c/2$.  (Here $\omega_c = e B / m c$ is the cyclotron frequency.)
In the strong magnetic field limit where $ \Omega/\omega_c \ll 1$
only the states in the lowest Landau level are relevant.
(Here $\omega_c = e B / m c$ is the cyclotron frequency.)
In the symmetric gauge the single particle states
 ($\phi_l(z) \sim z^l \exp( -z \bar z / 4 \ell^2)$)
in this level may be labeled by angular momentum
and have energy\refto{qdot}
$\epsilon_l = \hbar \omega_c/2 + \gamma (l+1)$ where $\gamma=m\Omega^2\ell^2$.
(Here $z = x + i y$ is the 2D electron
coordinate expressed as a complex number,
$\ell \equiv (\hbar c / e B)^{1/2}$ is the magnetic length, and the allowed
values of single particle angular momentum within the lowest Landau level
are $l=0,1,2,...$.)

Many-fermion states in the lowest Landau level may be described in a
boson language\refto{murray,xie}
which will prove useful in the fractional Hall regime.
The mapping between boson
and fermion descriptions can be understood in terms of a useful
property of many-fermion wavefunctions within the lowest
Landau level.  Any many-fermion wavefunction in this level
may be expressed in the form
$$
\Phi[z] = P(z_1,z_2,...,z_N) \prod_k \exp (- z_k \bar z_k / 4\ell^2)
\eqno(1)
$$
where $P[z]$ is an antisymmetric polynomial.  Since $P[z]$ must vanish
whenever any two coordinates are identical it follows that
$$
P[z] = \prod_{i<j \le N} (z_i-z_j) Q(z_1,z_2,...,z_N)
\eqno(2)
$$
where $Q[z]$ is a symmetric polynomial.  The fractional
quantum Hall effect is related to the existence of classes
of low-energy many-body states where the probability amplitude to
find any pair of electrons in a state of relative-angular-momentum less than
$m$ is identically zero.
States in these classes have wavefunctions of the form
$$
P[z] = \prod_{i<j \le N} (z_i-z_j)^m Q(z_1,z_2,...,z_N)
\eqno(2a)
$$
In the bulk such states exist only\refto{book,murray,macd}
when the Landau level
filling factor $ \nu \le 1/m $ and the chemical potential
jumps with increasing density at $\nu = 1/m$ as a
consequence.   In hard-core models\refto{haldanehc,kivtrug}
electrons interact only when their relative-angular-momentum
is less than $1/m$ so that the interaction energy is
identically zero in the corresponding class of low-energy states.
For a quantum dot with average $\nu < 1/m$ only these low-energy
states are relevant at temperatures low compared to the
fractional Hall gap and it is useful to consider a model where
the high-energy states are excluded from the Hilbert space.
(The relationship between hard-core models and
realistic models is discussed in more detail below.)
The products $Q[z] \prod_k \exp(-z_k \bar z_k/4\ell^2)$
are wavefunctions for $N$ bosons occupying states with $l\geq0$.
Since the low-lying excitations are completely described in terms
of the set of symmetric polynomials $Q$,
it follows that for $\nu < 1/m$, the number of low-energy
$N$-fermion wavefunctions at
total angular momentum $M_F$ equals the number of independent
$N$-boson wavefunctions, $g(N,M_B)$, at total angular momentum
$$
M_B = M_F - m N(N-1)/2.
\eqno(3)
$$
As we show below many properties of an ideal parabolic quantum
dot in the integer or fractional quantum Hall regimes can be expressed
in terms of $g(N,M_B)$.

We evaluate $g(N,M_B)$ using an iterative procedure in which
single-particle angular momenta are added incrementally to the
boson Hilbert space.  Define $g_k(N,M_B)$ to be the value of
$g(N,M_B)$ when only single-particle angular momenta from 0 to $k$
are included in the boson Hilbert space.  $g_0(N,M) = \delta_{M,0}$
since there is only one boson state for any number of electrons and it
has total angular momentum equal to zero.  When an additional single
particle state is added to the boson Hilbert space we have the
possibility of having any number of bosons in that state:
$$
g_k(N,M) = \sum_{N'=0}^N g_{k-1}(N-N',M-kN')
\eqno(4)
$$
(Note that $g_k(N,M) \equiv 0$ for $M <0$.)
For $k \ge M$, $g_k(N,M) = g(N,M)$ is independent of $k$.
Also, for $N \ge M$, $g_k(N,M) \equiv g_k(M)$ is independent of $N$.
Combining these, for $k,N\ge M$, $g_k(N,M)=g(M)$.
In \Fig{1} we plot
$s(N,M) \equiv \ln g(N,M)$ against $M^{1/2}$ for several
values of $N$.   For $N\geq M$,
the total number of states of boson angular momentum $M$
can also be identified
from the coefficients of the Taylor series expansion for the
boson partition function\refto{comment3}
$$
Z^B_k(\beta \gamma) = \prod_{k'=1}^k
 { 1 \over 1 - \exp( -\beta \gamma k')}
= \sum_{M=0}^{\infty} g_k(M) \exp ( - \beta \gamma M) .
\eqno(5)
$$
As shown in \Fig{1}, for large $M$
$s(M) \sim \sqrt{2/3} \pi M^{1/2} \sim 2.5650997 M^{1/2}$
with logarithmic corrections.  This result can be
obtained by comparing the mean values of the total boson
momentum calculated from the two forms for the right-hand
side of \Eq{5} in the limit of large $k$ and high
temperatures.  In this limit, $\ln Z^B \rightarrow \pi^2/(6\beta\gamma)$.

  To evaluate thermodynamic properties of the quantum dot
we need only note that the
energy of a fermion state in the class specified by \Eq{2a}
is determined by the total number
of electrons and the total angular momentum:
$$
E(N,M_F) =\gamma (N+M_F) = \gamma [N+ m N (N-1) /2 +M_B].
\eqno(6)
$$
It follows that the grand partition function for the dot
is given by
$$
Z_G = \sum_{N=0}^{\infty} \sum_{M_B=0}^{\infty} P(N,M_B),
\eqno(7a)
$$
where
$$
P(N, M_B) = g(N,M_B) \exp \left(\beta \gamma [N N_{\phi} -
m N (N-1)/2 -N]\right)  \exp ( - \beta \gamma M_B),
\eqno(7b)
$$
$\beta = 1/(k_B T)$, and
$$
N_{\phi} =( \mu - \hbar \omega_c/2 )/ \gamma =  B \pi R^2 / \Phi_0.
\eqno(8)
$$
$N_{\phi}$ is the number of flux quanta passing through the nominal
area of the dot.  Thermodynamic mean values of any quantity $f$
expressible in terms of $M_F$ or $N$ are now easily evaluated:
$$
\langle f(N,M_F) \rangle = Z_G^{-1} \sum_{N=0}^{\infty}
\sum_{M_B=0}^{\infty} f[N,M_B+ m N (N-1) /2] P(N,M_B) .
\eqno(9)
$$
In \Fig{2} we show plots of the $N_{\phi}$ dependence\refto{comment6}
of the magnetization of the dot for $m=1$.
We remark that in this system the magnetization
$M = \langle E \rangle / B $.\refto{comment4}
Note that as expected the magnetization has a component which is
periodic in $N_{\phi}$ with period one.
The results for $m=1$ are closely
related to those for non-interacting electrons
discussed previously by Sivan and Imry\refto{sivan}
and are precisely equivalent to
those trivially obtained by treating the system as a free
fermion system.
The equivalence of the present Bose
description and the Fermi description is closely related
to the bosonization transformations\refto{haldane} which
permit the exact treatment of models of a one-dimensional
electron gas. (See below.) The utility of
the Bose description is in treating the case $m \ne 1$,
where the system is {\it not} equivalent
to a free Fermi gas.  In \Fig{3} the
mesoscopic magnetization oscillations
are shown for the case of a fractional
Hall quantum dot with $m=3$.  Note that in this case
the magnetization has a component periodic in
$N_{\phi}$ with period three.

We can derive an analytic expression for the magnetization
which is valid for $ N_{\phi} \gg 1$.
In this limit $g(N,M_B) = g(M_B)$ for all
$M_B$ contributing importantly to the partition function.
Then the grand partition function factors into a boson
part which describes all neutral excitations of the system
and a part which describes the charged excitations:
$$
Z_G = Z_B \exp ( -\beta \Omega^*) \sum_{N=0}^{\infty} f(N+1/2),
\eqno(9)
$$
where $f(x) = \exp [ - \beta \gamma m (x-x^*)^2 /2]$,
$x^* = N_{\phi}/m +1 -1/m$, and
$ \Omega^* = - \gamma [N_{\phi}^2 - 2 N_{\phi} (1-m/2) + (1-m/2)^2]/2m$.
The boson excitations represent charge density oscillations
at the edge of the dot\refto{haldane,stone}.  The sum
over particle numbers can be performed using the Poisson
summation formula with the result
$$
Z_G = Z_B \exp ( -\beta \Omega^*) ( {2 \pi \over
 \beta \gamma m})^{1/2} \left[ 1 + 2
\sum_{s=1}^{\infty} (-)^s \exp ( - 2 \pi^2 s^2 / \beta \gamma
m ) \cos ( 2 \pi (N_{\phi} -1 ) s / m) \right].
\eqno(10)
$$
Except at very low temperatures the first term inside square
brackets will dominate.  An elementary calculation then leads
to the following expression for the magnetization:
$$
M = { \gamma \over B} \left[ { \pi^2 \over
6 (\beta \gamma)^2 } + {N_{\phi}^2 \over 2 m} -
{(1 - m/2)^2 \over 2m} + { 1 \over 2 \beta \gamma} \right]
+M_{osc}
\eqno(11)
$$
where
$$
M_{osc} = {\gamma \over B} \left[
{4 \pi N_{\phi} \over m \beta \gamma } \sin ( 2 \pi (N_{\phi}
-1) /m) + { 4 \pi^2 \over  m (\beta \gamma)^2 } \cos ( 2 \pi
(N_{\phi} -1) /m ) \right] \exp ( - 2 \pi^2 / \beta \gamma m)
\eqno(12)
$$
This expression for the magnetization agrees well with the numerical
results shown in \Fig{2} and \Fig{3}.

For a given chemical potential and parabolic confinement the
ground state is determined by a competition between the confinement
energy, which favors ground states of higher density (low total
angular momentum) and the interaction energy which favors ground
states of lower density (higher total angular momentum.)  This competition
has been discussed on the basis of exact diagonalization studies for
extremely small quantum dots by Maksym and Chakraborty\refto{maksym}.
For fixed chemical potential, the area of the dot
$2\pi(\mu-\hbar\omega_c/2)/(m\Omega^2)$ will decrease
with magnetic field, hence decreasing the strength
of the confinement.  The ground state for fixed electron number will
jump between different incompressible states.  For the hard-core model
the ground state will be the $m=3$ Laughlin state whenever the
confinement potential at the edge of the dot is weak compared to the
hard core potential strength.\refto{unpublished}.
For realistic
Coulomb interactions the dot density is
determined primarily, except at
extremely small dot sizes, by a competition between the Hartree
interactions in the dot and the confining potential.  The average
filling factor in the dot at strong fields will be approximately
the lesser of one and\refto{unpublished}
$$
\nu = \langle N\rangle^{1/3} \left({ 3 \pi \sqrt{2} \over 8} {\gamma \ell \over
 e^2 }\right)^{2/3}.
\eqno(12a)
$$
When correlations which favor particular densities are taken
into account the ground state for fixed $N$ will be
well-approximated by the finite $N$ Laughlin wavefunctions over
finite ranges of magnetic field
strength when the average filling factor is near $1/m$.  We
expect that
the results shown in \Fig{3} will apply for realistic interactions when
the average filling factor is near $1/3$.  Experimentally determined
values\refto{kastner} of the
curvature of the parabolic potentials give values of
$\gamma / k_B $ in excess of $0.1 K$ so that the effects shown in
\Fig{2} and \Fig{3} should be visible at accessible temperatures.
Similar mesoscopic oscillations occur in all
thermodynamic properties\refto{unpublished}
of the quantum dots and presumably in all physical properties.
For example we may expect that the conductance of the dot,
which is more easily measured than thermodynamic properties of
the dot, is closely related\refto{kinaret} to the particle number
fluctuations in the dot.

The oscillations shown in \Fig{3} reflect a fundamental
property of fractional Hall states.  Near $\nu = 1/3$ the
properties of the system in contact with a particle reservoir
have contributions which are periodic in $N_{\phi}$ with
period three.  These oscillations express the fact that on
average one electron is added to the system for each three
additional flux quanta.  Hence they can be observed only
when the field dependence is studied at fixed chemical potential
rather than fixed particle number.
The quantized Hall conductance is
proportional to the charge added to the system per
flux quantum in the thermodynamic limit, {\it e.g.\/}, $e/3$ for $\nu=1/3$.
To obtain this result using an effective theory of non-interacting
fermions\refto{wen} would require the fermions to have a
charge $e/\sqrt{3}$.   Such a theory would
would give mesoscopic oscillations with an incorrect period.

This work was supported in part by the National Science
Foundation under grant DMR-9113911.  AHM acknowledges
helpful conversations with Ian Affleck, M. A. Kastner,
Daniela Pfannkuche and
Xiao-Gang Wen and the hospitality of the University of
New South Wales where part of this work was completed.

\vfill\eject

\references

(a) Permanent address: Department of Physics, Indiana University,
Bloomington IN, 47401.

\refis{kinaret} J.M. Kinaret, Y. Meir, N.S. Wingreen, P. Lee, and X.G. Wen,
\prb 45, 9489, 1992.

\refis{unpublished} A.H. MacDonald and M.D. Johnson, unpublished.

\refis{maksym} P. A. Maksym and Tapash Chakraborty, \prl 65, 108, 1990.

\refis{comment6} $ B d N_{\phi} / d B  = N_{\phi}- \hbar
\omega _c / 2 \gamma$.
In the strong magnetic field limit
the number of flux quanta through a dot decreases with field
because kinetic energy quantization depopulates the dot.

\refis{book} See for example, A. H. MacDonald,
{\it A Perspective on the Quantum Hall Effect},
(Jaca Books, Milano, 1989).

\refis{comment4} We give the magnetization and energy with respect to
the values for free electrons in the lowest Landau level,
$M_{free} = - \mu_B N$.

\refis{xie} X.C. Xie, Song He, and S. Das Sarma, \prl 66, 389, 1991.

\refis{murray} A.H. MacDonald and D.B. Murray, \prb 32, 2707, 1985.

\refis{macd} A.H. MacDonald, \prl 64, 222, 1990.

\refis{stone} Michael Stone, \prb 42, 8299, 1990; {\it Ann. Phys. (NY)\/}
{\bf 207}, 38 (1991); Michael Stone, H.W. Wyld, and R.L. Schult,
\prb 45, 14156, 1992.

\refis{wen} X.G. Wen, review article submitted to {\it Int. J. Mod. Phys. B\/},
and references therein.

\refis{kivtrug} S.A. Trugman and S.A. Kivelson,
Phys. Rev. B {\bf 31}, 5280, (1985).

\refis{laugh} R.B. Laughlin, Phys. Rev. Lett., {\bf 50}, 1395 (1983).

\refis{haldanehc} F.D.M. Haldane, {\it The Quantum Hall Effect},
 R.E. Prange and S.M. Girvin eds.,
 (Springer Verlag, New York, 1986), Chapter 8.

\refis{haldane} F.D.M. Haldane, J. Phys. C {\bf 14}, 1585 (1981).

\refis{sivan} U. Sivan and Y. Imry, \prl 61, 1001, 1988.

\refis{comment3} This is the boson grand partition function for
bosons occupying states $k'=1$ to $k$ with the chemical
potential set to the energy of the $k=0$ state.
This partition function accounts for the
neutral phonon-like collective excitations of the system.
For $N > M $ the neutral excitations and charged excitations
of the system are independent so that the partition function may
be factored for large systems or low temperatures (see text).

\refis{hansen} W. Hansen, T.P. Smith III, K.Y. Lee, J.A. Brum,
C.M. Knoedler, J.M. Hong, and D.P. Kern, \prl 62, 2168, 1989.

\refis{qdot}  For recent reviews see
U. Merkt, Advances in Solid State Physics, {\bf
30}, 77 (1990); Tapash Chakraborty,
Comments on Condensed Matter Physics {\bf 16}, 35 (1992);
M.A. Kastner, Rev. Mod. Phys. {\bf 64}, 849 (1992).

\refis{kastner} P.L. McEuen, E.B. Foxman, U. Meirav,
M.A. Kastner, Y. Meir, Ned S. Wingreen, and S.J. Wind,
\prl 66, 1926, 1991; P.L. McEuen, E.B. Foxman, Jari Kinaret,
U. Meirav, M.A. Kastner, Ned S. Wingreen, and S.J. Wind,
\prb 45, 11419, 1992.

\refis{qhe} K. von Klitzing, G. Dorda, and M. Pepper,
\prb 45, 494, 1980; D.C. Tsui, H.L. Stormer, and A.C. Gossard,
\prl 48, 1559, 1982.

\endreferences

\vfill
\eject

\figurecaptions

\figis{1} $s(N,M) \equiv \ln g(N,M)$ {\it vs\/}. $M^{1/2}$. The
larger `entropy' is for $N = \infty$ and the smaller
is for $N=10$.  The dashed line is $\sqrt{2/3}\pi M^{1/2}$.

\figis{2} Magnetic oscillations of non-interacting
electrons in a parabolic quantum dot at
$\beta \gamma =10$
and at $\beta \gamma =1$.  (The curve with larger oscillations
is for lower temperature.)
The plotted quantity is $M + \mu_B N - (\gamma / 2 B) ( N_{\phi}^2 -1/4)$.
These results are valid for $\beta \hbar \omega_c \gg 1$.

\figis{3} Magnetic oscillations of electrons with
hard-core interactions in a parabolic quantum dot with
$\nu < 1/3$ at
$\beta \gamma =10$ and at $\beta \gamma =1$.
The plotted quantity is $M + \mu_B N - (\gamma / 6 B) ( N_{\phi}^2 -1/4)$.
These results are
valid for $\beta V \gg 1$ where $V$ is the strength of the
hard core interaction.

\endfigurecaptions
\end